# Thermoelectric Spin-Transfer Torque MRAM with Sub-Nanosecond Bi-Directional Writing using Magnonic Current

Niladri N. Mojumder[1,2], Kaushik Roy[1] and David W. Abraham[2]

[1] *Department of Electrical & Computer Engineering, Purdue University, West Lafayette, IN 47907, USA*

[2] *IBM T. J. Watson Research Center, Yorktown Heights, NY 10598, USA*

*Abstract -* A new genre of Spin-Transfer Torque (STT) MRAM is proposed, in which bi-directional writing is achieved using thermoelectrically controlled magnonic current as an alternative to conventional electric current. The device uses a magnetic tunnel junction (MTJ), which is adjacent to a non-magnetic metallic and a ferrite film. This film stack is heated or cooled by a Peltier element which creates a bi-directional magnonic pulse in the ferrite film. Conversion of magnons to spin current occurs at the ferrite-metal interface, and the resulting spin-transfer torque is used to achieve sub-nanosecond precessional switching of the ferromagnetic free layer in the MTJ. Compared to electric current driven STT-MRAM with perpendicular magnetic anisotropy (PMA), thermoelectric STT-MRAM reduces the overall magnetization switching energy by more than 40% for nano-second switching, combined with a write error rate (WER) of less than $10^{-9}$ and a lifetime of 10 years or higher. The combination of higher thermal activation energy, sub-nanosecond read/write speed, improved tunneling magneto-resistance (TMR) and tunnel barrier reliability make thermoelectric STT-MRAM a promising choice for future non-volatile memory applications.

Over the last few decades, the trend towards increased adoption of embedded memory to increase the bandwidth of high performance processors and mobile system-on-chips (SoCs), has prompted research in novel memory technologies [1-4]. Among them, spin-transfer torque magnetic random access memory (STT-MRAM) has stimulated significant interest due to its unique combination of properties, including data non-volatility, unlimited endurance, negligible static leakage power, high performance and high integration densities [5-7]. Unlike magnetic field-driven MRAM [4], STT-MRAM offers lower switching current, simpler cell architecture, reduced manufacturing cost and excellent scalability (with a bit-cell size as small as $6F^2$), making it a competitive choice for future technology nodes [6-8].

An electric current passing through a pair of ferromagnetic electrodes separated by a metallic spacer or tunnel barrier exerts pseudo-torque on the magnetic moment of the individual electrodes [9-10]. The magnitude of this spin-transfer torque is proportional to the electric current [9-12]. Switching using this torque requires a pulse length of at least 5-10 ns [13-14], unless the device is overdriven with currents significantly higher than the switching threshold. For applications to MRAM, such high currents are incompatible with the requirement to avoid breakdown of the tunnel barrier during its lifetime (say, 10 years). For slower switching speeds, breakdown of the barriers is typically avoided by limiting the write voltage across the barrier to around 400 mV and using a tunnel barrier with resistance-area (RA) product in the range of $5 - 10$ $\Omega-\mu m^2$ [15]. These lower RA barriers are more susceptible to breakdown and have lower tunneling magneto-resistance (TMR), which reduces the read margin. Lowering the switching threshold (for example by making the free layer thinner) would reduce the activation energy, causing poor retention due to thermally activated switching. Hence, it is difficult to achieve fast sub-nanosecond switching of the free layer using spin-transfer torque, without increasing errors due to tunnel barrier breakdown, read, and thermal instabilities.

Recently, Slonczewski proposed initiating spin-transfer torque by heat flux flowing through the free layer and an insulating reference ferrite [16]. This torque is generated by the creation or annihilation of magnons in the ferrite and subsequent conversion to electron spin current at the interface between the ferrite and a non-magnetic metal film. Depending on the direction of heat flow, this spin-transfer torque tends to align or anti-align the free layer magnetic moment with the ferrite magnetic moment. With proper suppression of heat energy carried by phonons inside the ferrite [17-18], the proposed magnonic spin-transfer torque has a quantum yield of almost two orders of magnitude higher than achievable using conventional electric current through the magnetic tunnel junction (MTJ) [16]. This enhanced quantum yield opens up the possibility of designing new spintronic devices with low power and high speed of operation. However, since the sign of the thermagnonic spin-transfer torque depends on the direction of heat current, magnonic spin-transfer initiated by Joule heating of a resistor adjacent to the MTJ is not directly applicable to MRAM, where bi-directional switching of the free layer is required.

In this article, we propose an alternative MRAM design in which the storage layer is efficiently switched by magnonically generated spin-transfer torque. An insulating ferrite film with uniaxial magnetic anisotropy collinear to that of the nearby free layer is used to flip the free layer moment bi-directionally during write. The free layer can be switched to be either parallel or anti-parallel to the ferrite orientation, depending principally on the direction of heat current flowing through the insulating ferrite and the free layer separated by a thin metal spacer [16]. By using a Peltier element in the device stack, it is possible to simply reverse the direction of heat flow, and hence, obtain bidirectional writing. Conversion from magnons to electrical spins occurs at the ferrite-metal spacer boundary.

Figure 1 shows the schematic of the proposed device configuration to utilize thermoelectrically controlled magnonic spin-momentum transfer for bi-directional magnetic switching. The uniaxial magnetic anisotropy of the ferrite film is perpendicular to the x-y plane and collinear to those of reference and free magnetic layers [all along z-direction]. For this structure, we estimate the thermoelectrically initiated magnonic spin-transfer torque by considering the exchange coupling between the ferrite moment (***M**$_{Ferrite}$*) and conductive s-electrons in the adjacent metal spacer through interfacial atomic monolayer with paramagnetic 3d-electron-spin moments ($\sum$) [16, 19]. In Slonczewski's model, ***σ*** and ***F*** are respectively the average thermal moment of the s-electron-spins per unit area of the metal spacer and an effective molecular field at the ferrite-metal interface. $M_{Ref}$ and $m_{Free}$ are respectively the magnetic moments of the ferromagnetic pinned and free layers, separated by a thin tunnel barrier (typically MgO). The transient response of $\sigma$ can be understood by considering its coupling to the monolayer moment $\Sigma$ and solving the Bloch equation appropriately [16, 20].

A temperature differential δT between the ferrite spins including the magnetic monolayer and the adjacent non-magnetic metal develops an interfacial heat current. A steady flow of heat current from the heat source to the thermal sink yields a non-vanishing time rate of $\sigma$, defined as the thermagnonic spin-transfer torque per unit free layer area. In practice, the thermal conductance due to interfacial phonon scattering, and ferrite-to-monolayer and monolayer-to-metal heat transfer efficiencies jointly determine the effective spin-momentum transfer due to ferrite magnons [21-22]. The magnitude of the thermagnonic spin-transfer torque along z-direction (perpendicular to the film plane, x-y) can be estimated as [16]:

$$|\tau_z| = \frac{\pi.S.(S+1).N_d.(J_{sd}\rho)^2.F(T)}{3.\hbar.T}|\delta T| \qquad (1)$$

where *S* is the spin quantum number of the paramagnetic metallic atoms in the magnetic monolayer, $N_d$ is the number of magnetic ions or atoms per unit area of the magnetic monolayer, *F(T)* is the molecular-field exchange splitting of the magnetic ions sitting at the ferrite-metal interface at temperature T, *δT* is the effective temperature differential across the ferrite-metal interface with proper sign dependent on the direction of heat flow, $J_{sd}$ is the on-site s-to-d-electron exchange coupling, ρ is the s-electron density per atom, respectively, and $\hbar$ is the Dirac constant.

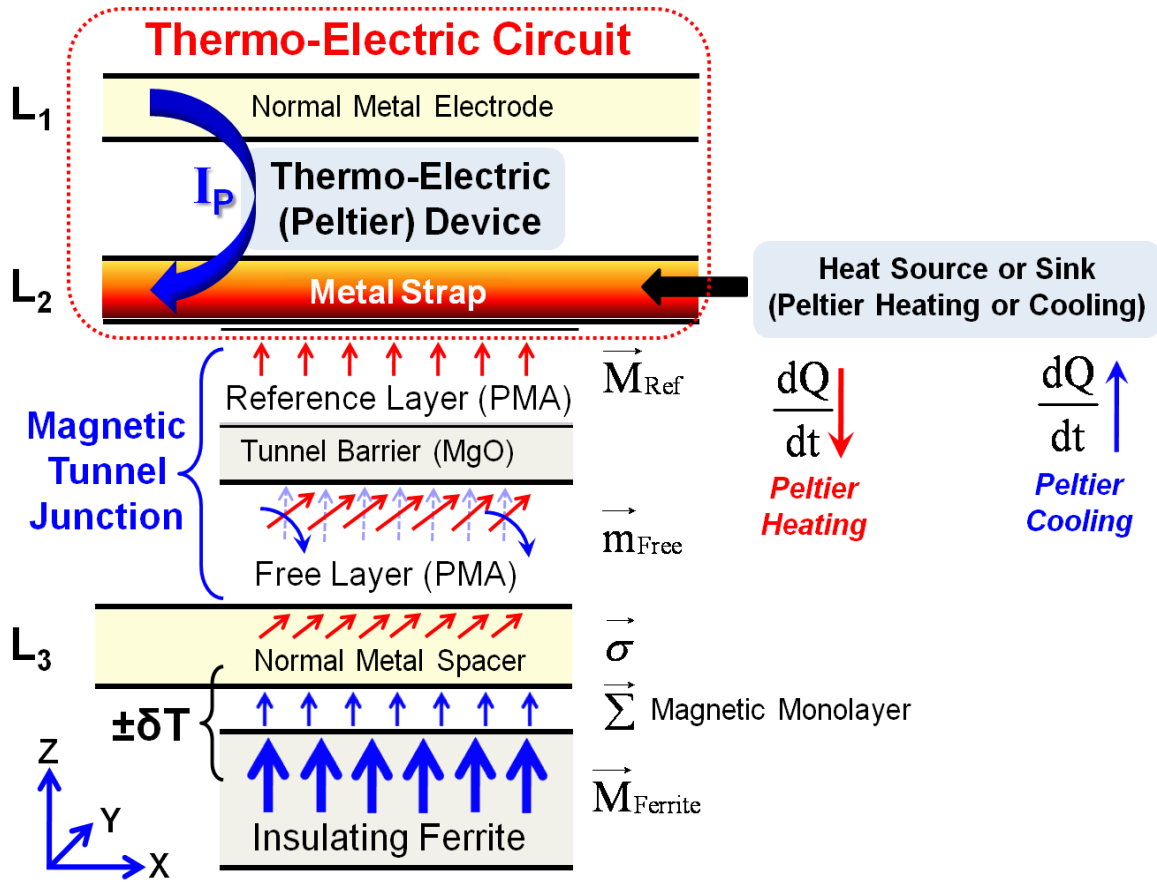

**Fig. 1** Schematic of the proposed thermoelectric STT-MRAM device using bipolar thermagnonic current for bi-directional free layer switching

It is worth noting that the sign of the perpendicular magnonic torque $\tau_z$ depends on the direction of heat current flow across the ferrite-metal interface. In the model proposed in Fig. 1, a positive δT (equal to $T_{Ferrite}$ - $T_{MS}$) tends to align free layer moment with *$M_{Ferrite}$*. A negative δT anti-aligns *$m_{Free}$* with *$M_{Ferrite}$*. In our proposed device, the sign and magnitude of δT is controlled using thermo-electric Peltier effect, thus giving the ability to achieve fast nanosecond bi-directional switching.

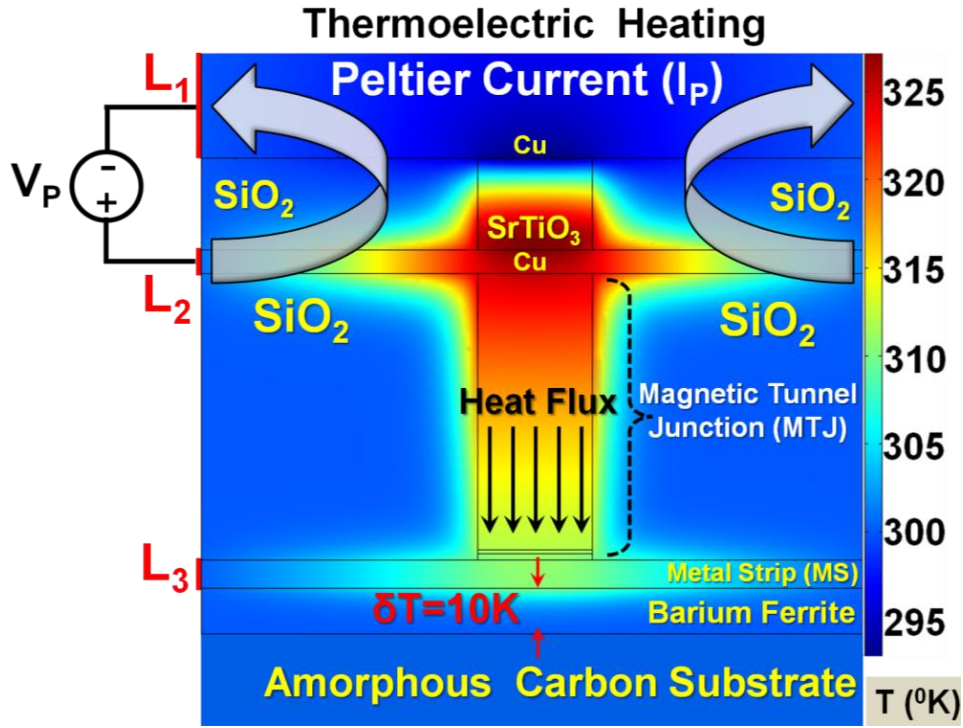

**Fig. 2(a)** Steady state temperature distribution in the thermoelectric STT-MRAM device under Peltier heating, enhanced by joule heating. The metal strip (MS) $L_3$ is at a higher temperature than the Barium Ferrite. $\delta T > 0K$

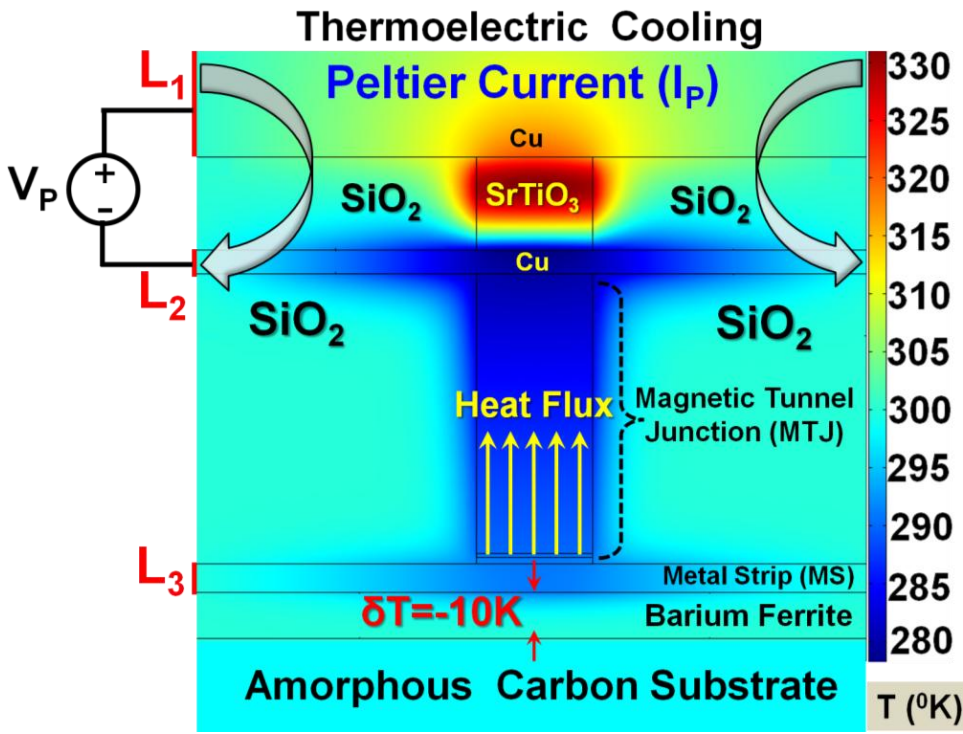

**Fig. 2(b)** Steady state temperature distribution in the thermoelectric STT-MRAM device under Peltier cooling, hindered by joule heating. The metal strip (MS) $L_3$ is at a lower temperature than the Barium Ferrite. $\delta T < 0K$

The detailed device structure of the proposed thermoelectric STT-MRAM using magnonic current for bi-directional free layer switching is shown in Fig. 2. For the magnetic tunnel junction, we use the vertical stack structure as described and fabricated in [23]. The MTJ with PMA is placed vertically above a non-magnetic metallic spacer, adjacent to a ferrite film grown on amorphous carbon substrate. A thin layer of peltier material, such as Nb-doped Strontium Titanate, sandwiched between two metal straps $L_2$ and $L_3$ is placed right above the magnetic tunnel junction. .   The composite stack forms a thermoelectric circuit, generating bidirectional magnonic current through the free magnetic layer [24]. A bipolar electric potential between $L_1$ and $L_2$ sends a current $I_P$ through the Peltier element. On assertion of a positive potential at $L_2$ relative to $L_1$, the metal strap $L_2$ heats, converting it to a heat source as designated in Fig. 1. Conversely, a negative potential at $L_2$ relative to $L_1$ cools down $L_2$, transforming it into a heat sink.  Changing the polarity of the applied voltage ($V_P$) between $L_1$ and $L_2$ thus develops a bi-directional heat flux flowing normally across the metal ($L_3$)-Barium Ferrite interface as shown in Fig. 1 and 2(a, b), and in turn creates a bipolar magnonic current essential for bi-directional magnetic switching of the free layer.  As we shall see, a magnonic current with higher spin-torque efficiency in combination with an efficiently designed Peltier circuit helps in scaling down the current $I_P$ in thermoelectric STT-MRAM. The design with thermoelectric circuit on top provides a broader choice of magnetic and electrode materials, since the ferrite (which is typically grown at high temperatures) is grown first, below the magnetic and Peltier stack [25].

In thermoelectric STT-MRAM, the Peltier current $I_P$ required for heating/cooling of the metal strap $L_2$ during write essentially determines the overall switching energy dissipation.  To estimate the electro-thermal efficiency of the proposed device structure (Fig. 1), we solve for transient thermoelectric conduction coupled with Landau-Lifshitz-Gilbert-Slonczewski (LLGS) equation in a single domain magnetic landscape [16, 26-27].  First, we solve the Fourier heat transfer equation [28] coupled with Poisson's equation self-consistently in field variables temperature 'T' and electric potential 'V', using a commercial finite element analysis  package [29-30]. For all subsequent micro-thermal analysis of the proposed device structure shown in Fig. 2, the geometry parameters are chosen as follows: Individual layer thicknesses are amorphous carbon(100nm) /BaFe (20nm)/ Cu (10nm)/ free layer (3nm)/ MgO (1.2nm)/ reference layer (6nm)/ PtMn (25nm)/ Ta (60nm)/ Cu(10nm)/ SrTiO$_3$(40nm)/Cu(50nm), and the diameter of the circular MTJ pillar is 50nm, and the Peltier current I$_P$ is described by a rectangular pulse of amplitude V$_P$ and pulse width PW of 0.15ns. Fig. 3 shows the transient response of the temperature difference δT, across the ferrite-metal interface. The thermal time constant of the proposed nanostructure causes a gradual decay in δT as predicted in Fig. 3.

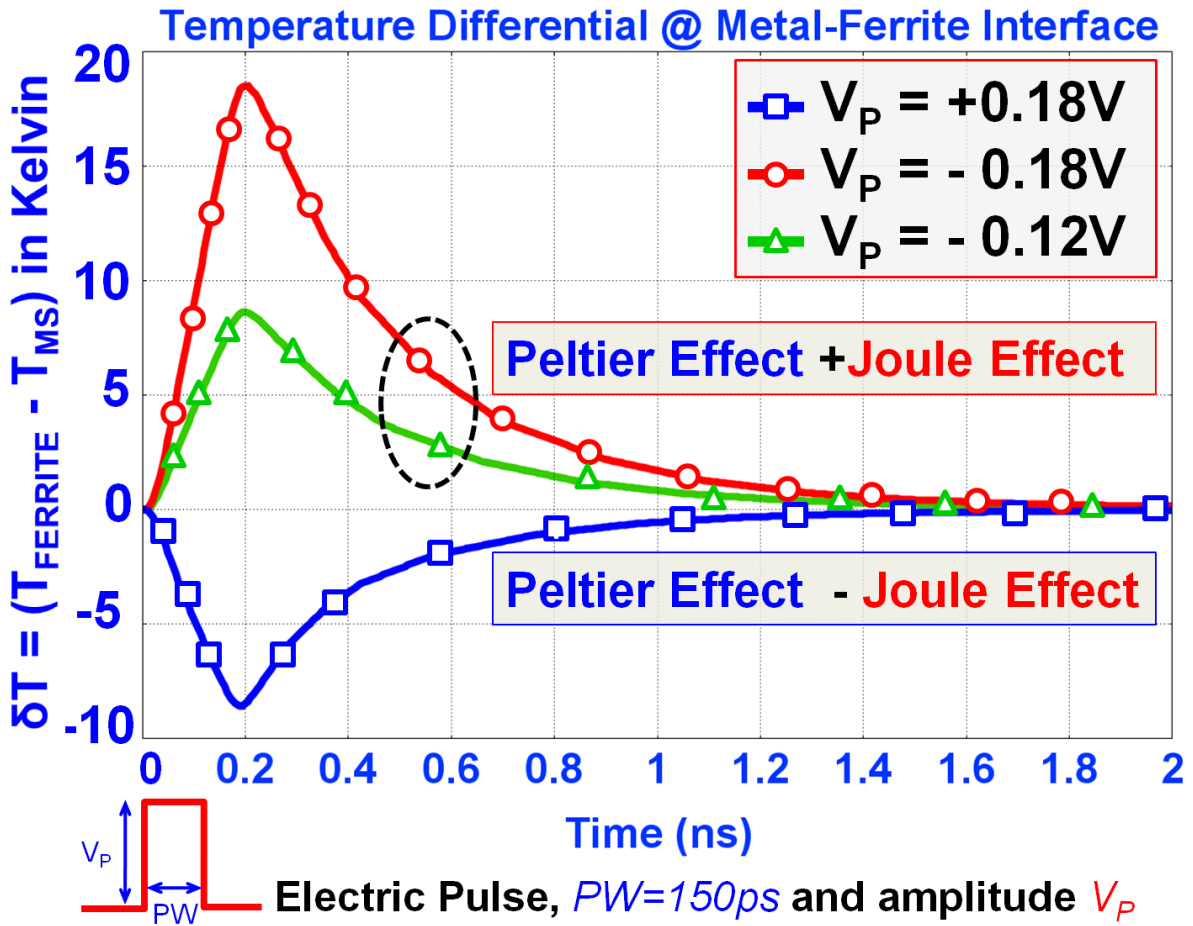

**Fig. 3** Transient response of heating/cooling in thermoelectric STT-MRAM under short duration Peltier current (I$_P$) pulse; The thermal time constant of the system ensures gradual decay of the temperature profile from |δT$_{MAX}$|, achieved using a short duration voltage pulse of magnitude V$_P$ and pulse width PW.

In a recent communication [31], the time-resolved heat-flow dynamics in ferromagnetic thin films have been studied experimentally. A large magnonic contribution to the total heat current flowing normal to the film surface has been reported over a broad temperature range. Subject to a pulsed heating/cooling, the local magnon temperature rises/diminishes instantaneously due to fast (~ps) thermal energy transfer with phonons. Once the required temperature differential is developed across the ferrite-metal interface, magnonic spin-transfer torque acts instantaneously on the storage layer to switch its magnetization.

Assuming a transient $\delta T$ profile as predicted in Fig. 3, we solve LLGS equation in the presence of thermally induced stochastic magnetic noise, with thermagnonic spin-transfer torque terms included, as modeled in Eq. 1. Subject to a maximum temperature differential $\delta T_{MAX}$ of ±8K across the metal strip $L_3$ and ferrite interface, the P-AP and AP-P switching characteristics are shown in Fig. 4. Sub-nanosecond switching delay is achieved, with no electric current flowing through the thin tunnel barrier. In an electric current driven STT-MRAM with PMA, the critical switching current density $J_E$ for a delay of 1ns and with a switching failure probability of $10^{-9}$ has been estimated to be close to 14 MA/cm$^2$ at 300K [32]. In order to avoid voltage-induced breakdown, the RA product of the tunnel barrier would have to be reduced to well below 5 $\Omega$-$\mu$m$^2$. In practice it has been difficult to achieve such low RA barriers which are pinhole-free [33]. Thus, in conventional STT-MRAM, sub-nanosecond precessional switching is difficult to achieve in combination with a lifetime of ~10 years, as the tunnel barrier breaks down due to the excessive current requirement for switching [15]. To eliminate the stochastic thermal incubation delay during free layer switching, we assume an angular tilt ($\theta_{Tilt}=25^0$) in ferrite magnetic anisotropy relative to that of the free layer [34-36]. The non-collinear alignment of the ferrite and free layer anisotropies ensure the reproducible (error-free) switching of the free layer magnetization under thermagnonic spin-transfer torque. The proposed thermoelectric STT-MRAM offers sub-nanosecond, low power and reproducible magnetic switching, with lifetime of 10 years or higher. Compared to electric current driven STT-MRAM with perpendicular magnetic anisotropy (PMA), thermoelectric STT-MRAM reduces the overall magnetization switching energy by more than 40% for nano-second switching, combined with an estimated Write Error Rate (WER) of less than $10^{-9}$.

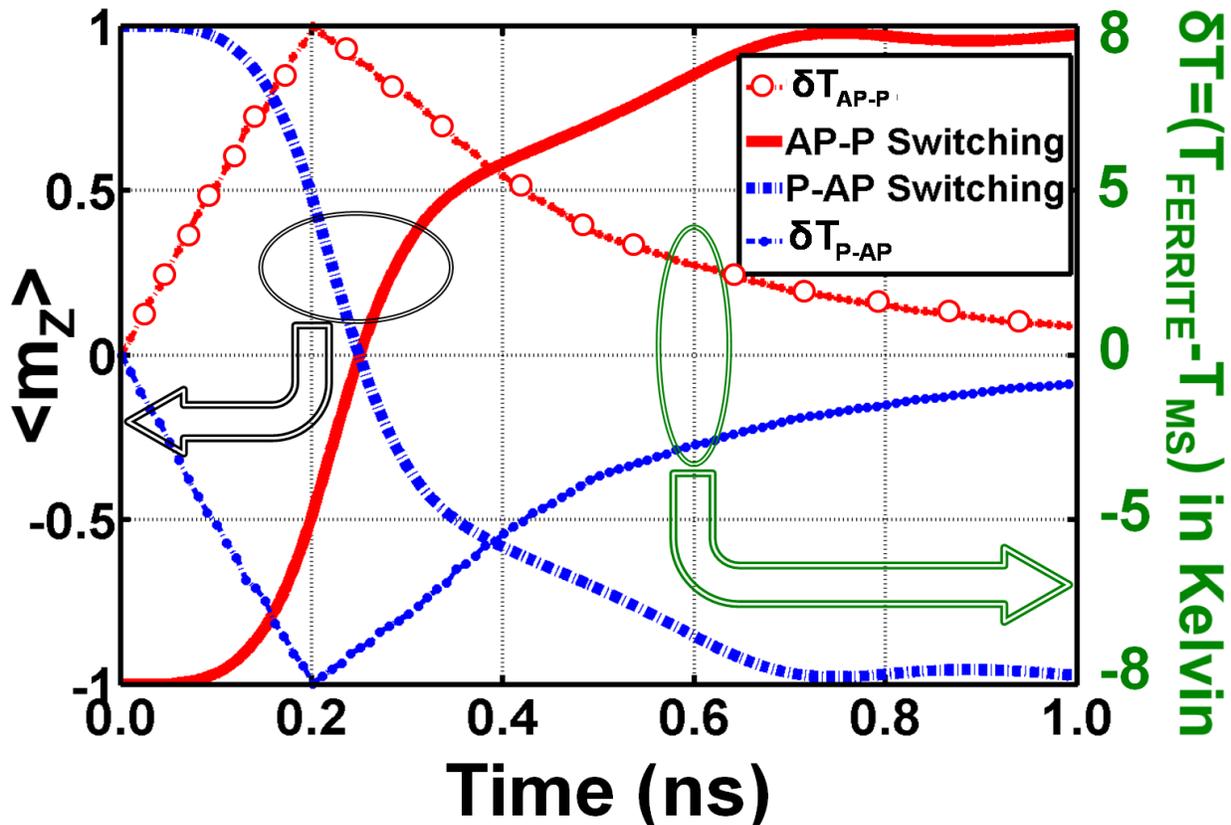

**Fig. 4** Bi-directional magnetic switching using bipolar thermagnonic current for $\delta T= \pm 8K$ with free layer anisotropy perpendicular to the film plane (along z-axis). The simulation parameters are chosen as: $E_a$=80kT, $M_{sat}$=850emu/cm$^3$, $H_{ku2}$=880emu/cm$^3$, vol=50x50x3nm$^3$, S= 2.5; $N_d$=4x10$^{18}$/m$^2$; interfacial spin moment in ferrite (FT)=30meV; T=300K; On-site sd exchange coupling ($J_{sd}$)=-0.5eV; Density of S-electrons per atom ($\rho$)=0.15 eV$^{-1}$.spin$^{-1}$.atom$^{-1}$;

In an STT-MRAM bit cell, the presence of an access device in series with the MTJ during read reduces the effective TMR of the bit cell, which can be expressed as:

$$\text{TMR}_{CELL} = \frac{\text{TMR}_{MTJ}}{(1+\frac{R_{ACCESS}}{R_P})} \qquad (2.a)$$

$$\text{TMR}_{MTJ} = \frac{R_{AP}-R_P}{R_P} \qquad (2.b)$$

$R_{ACCESS}$, $R_{AP}$ and $R_P$ are the resistance of the access device during read, and anti-parallel and parallel resistances of the MTJ, respectively. To ensure a high cell TMR during read, one needs to maximize both $R_P$ and $TMR_{MTJ}$ for a fixed width of the access transistor. This can be done by increasing the thickness of the tunnel barrier in the MTJ pillar. In a state-of-the-art STT-MRAM, higher switching current density electrically limits the use of a thicker tunnel barrier and $T_{OX}$ is typically kept below 1.2nm for a WER less than $10^{-9}$. In the proposed thermoelectric STT-MRAM, as no electric current is required to flow through the tunnel barrier, a thicker tunnel barrier ($T_{OX}$ ~ 1.5nm) can be incorporated. Higher RA product enhances the bit cell TMR and significantly reduces the disturb failure probabilities during read [12, 37]. Fig. 5(a-c) shows the effect of tunnel barrier thickness ($T_{OX}$) and voltage on $R_{AP}$, $R_P$, TMR and tunneling current density ($J_E$) of an MTJ with MgO as tunnel barrier [37-38]. As shown in Figs. 5(a) and (b), the differential resistances in parallel and anti-parallel states of an MTJ increases almost by an order, when $T_{OX}$ is changed from 1.2nm to 1.5nm. The thicker tunnel barrier provides an enhanced guard band against 'soft' oxide breakdown [15] and memory disturb failures, in addition to achieving almost 40% higher MTJ TMR than the conventional STT-MRAM.

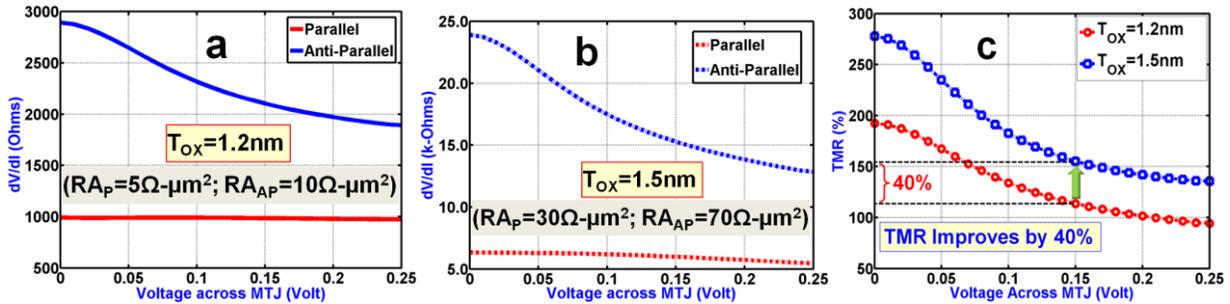

**Fig. 5** Bias-voltage dependence of parallel and anti-parallel differential resistances (dV/dI) of an MTJ with **(a)** $T_{OX}$=1.2nm and **(b)** $T_{OX}$=1.5nm. **(c)** A thicker tunnel barrier helps in boosting up the MTJ TMR.

In this article, we propose a new genre of STT-MRAM, using thermoelectrically controlled magnonic spin-transfer torque for low power, fast, reliable and error-free magnetic switching. A bipolar electric current through a thin film Peltier element generates a bidirectional magnonic current at a ferrite-metal interface placed in contact with a ferromagnetic free layer. Conversion of magnonic current into spin-transfer torque is used to achieve sub-nanosecond bidirectional switching of the free layer.

Compared to the conventional STT-MRAM with PMA, the proposed thermoelectric STT-MRAM reduces the magnetization switching energy by almost 40% for a switching delay of 1ns and WER less than $10^{-9}$. The combination of higher thermal activation energy, sub-nanosecond read/write speed, improved TMR, and tunnel barrier reliability make thermoelectric STT-MRAM a promising choice for future non-volatile memory applications.